\documentclass[aps,prd,twocolumn,superscriptaddress,floatfix,nofootinbib]{revtex4-1}

\usepackage[T1]{fontenc}

\usepackage{amsmath,amssymb,amsfonts}
\usepackage{graphicx}
\usepackage[normalem]{ulem}
\usepackage{subcaption}
\usepackage{placeins}
\usepackage[colorlinks=true,citecolor=blue,linkcolor=blue,urlcolor=blue]{hyperref}
\usepackage{cleveref}

\begin{document}


\title{Dissipative Losses In Black Hole-Induced Vacuum Decay}

\author{Michael Geller,}
\affiliation{School of Physics and Astronomy, Tel Aviv University, Tel-Aviv 69978, Israel}

\author{Ofri Telem}
\affiliation{Racah Institute of Physics, Hebrew University of Jerusalem, Jerusalem 91904, Israel}

\begin{abstract}
We address the long-standing puzzle of false vacuum decay catalyzed by black holes. Naively, small black holes with large Hawking temperatures can generate highly-boosted true vacuum bubbles in the early universe and trigger vacuum decay without any exponential suppression. Working in the thin-wall regime of the $\phi^4$ and sine-Gordon models, we show that radiative losses play a crucial role in decelerating these bubbles and preventing runaway vacuum decay. We find that while the production rate is enhanced compared with vacuum tunneling in some parts of the parameter space, it is always exponentially suppressed. 
\end{abstract}
\maketitle

\section{Introduction}
Can small black holes catalyze early-universe phase transitions? This question, first considered in \cite{Hiscock:1987hn,Berezin:1988ur,Berezin:1991rp,Arnold:1989cq,Dai:2019eei,Carr:2020gox}, has been the focus of considerable debate in the literature \cite{Hiscock:1987hn,Berezin:1988ur,Berezin:1991rp,Arnold:1989cq,Dai:2019eei,Carr:2020gox,Gregory:2013hja,Burda:2015yfa,Burda:2015isa,Burda:2016mou,Mukaida:2017bgd,Kohri:2017ybt,Hayashi:2020ocn,Shkerin:2021xsh,Shkerin:2021uun,Briaud:2022txw,Strumia:2022jil,Rossi:2025fix}. The detailed analysis \cite{Gregory:2013hja} concluded that primordial black holes (PBH) with masses up to $10^{9}M_{pl}$ may destabilize the vacuum prior to their evaporation. Subsequent works \cite{Burda:2015yfa,Burda:2015isa,Burda:2016mou,Mukaida:2017bgd,Kohri:2017ybt,Hayashi:2020ocn} considered this question in the context of the metastability of the Higgs potential, citing significant lower bounds on the early-universe number density of PBH \cite{Dai:2019eei,Carr:2020gox}.

The study of early universe phase transitions originated in the seminal works by Coleman et al. \cite{Coleman:1977py,Callan:1977pt,Coleman:1977th} and their subsequent gravitational refinement \cite{Coleman:1980aw}\footnote{For modern treatments of complex saddles, Picard--Lefschetz theory, and real-time path-integral tunneling, see
\cite{Witten2010QM,Witten2010CS,BasarDunneUnsal2013,CristoforettiEtAl2013,TanizakiKoike2014,ChermanUnsal2014,BehtashEtAl2017,AlexandruEtAl2016,HertzbergYamada2019,AiGarbrechtTamarit2019,MouSaffinTranbergWoodward2019,MouSaffinTranberg2019,NishimuraSakaiYosprakob2023,BlumRosner2023}}. Using semiclassical methods, Coleman considered the case of a scalar theory whose potential has nearly degenerate vacua. While the universe is initially in the higher (``false'') vacuum, bubbles of the lower (``true'') vacuum nucleate quantum mechanically. The evolution of these bubbles depends on a critical radius $r_c$, defined as the outer turning point of the effective potential. When bubbles are produced at rest beyond $r_c$, they can overcome their surface tension and experience runaway growth, ultimately catalyzing a phase transition into the true vacuum. In Coleman's original study, true vacuum bubbles are generated at zero energy. Consequently, their generation rate is exponentially suppressed by the need to tunnel through their surface tension barrier of height $E_{top}$ to the critical radius $r_c$ (see Fig.~\ref{fig:bubble_potential}). This is not the case for true vacuum bubbles produced around small black holes; indeed, due to the latter's large Hawking temperature, true vacuum bubbles could in principle emerge with a nonzero boost, potentially allowing them to traverse their surface-tension barrier \textit{classically}.

Naively, the effect of a black hole Hawking temperature implies a standard thermal $O(3)$ calculation of the Euclidean bounce. However, this is imprecise, as astrophysical black holes are not in thermal equilibrium with their surroundings \cite{Unruh:1976db}. Recent ``first-principles'' approaches \cite{Shkerin:2021xsh,Shkerin:2021uun,Briaud:2022txw} address this question by solving for a generalized bounce satisfying Unruh boundary conditions. These calculations have so far only been performed for toy models, with promising results.

In this work, we present a complementary approach. In the thin-wall approximation of the $\phi^4$ and sine-Gordon models, we show that the problem factorizes into distinct stages:
(i) Hawking production of boosted bubble walls in the near-horizon region;
(ii) a burst of radiative losses into scalar radiation, after which the bubble energy is capped at a scale $E_\oplus$ that depends on the black hole mass.
(iii) if $E_{\oplus}>E_{top}$, unsuppressed classical expansion, while if $E_{\oplus}<E_{top}$, Coleman-like tunneling through the barrier; and
(iv) runaway bubble expansion beyond the critical radius.

Although our analysis does not prove the absence of other production modes, it paints a clear physical picture of bubble production. In particular, we identify the burst of radiative energy loss from highly boosted bubbles after their production as the main limitation on black-hole induced vacuum decay. This radiation caps the boost of bubbles that could otherwise classically traverse the surface-tension barrier. As a result, bubble production is always exponentially suppressed, albeit with a mild enhancement in parts of the parameter space. 

\section{Black hole induced False-Vacuum Decay}
We consider a real scalar field in Lorentzian signature $(-,+,+,+)$, in the background of a Schwarzschild black hole,
\begin{eqnarray}\label{eq:4daction}
S=\int d^4x\sqrt{-g}\left[-\frac{1}{2}g^{\mu\nu}\partial_\mu \phi\partial_\nu \phi-V(\phi)\right] \, ,
\end{eqnarray} 
where the Schwarzschild metric is $ds^2=-f(r)\,dt^2+f^{-1}(r)\,dr^2+r^2d\Omega^2$, with $f(r)=1-r_s/r$ and $r_s$ the Schwarzschild radius. We assume that $V(\phi)$ has at least two approximately degenerate minima at $\phi=\phi_\pm$. As in Coleman's original setting, we assume that the universe is in a metastable (``false'') vacuum $\phi_+$, while the global (``true'') vacuum is $\phi_-$. In this case, the theory admits bubble solutions, interpolating between the true vacuum $\phi_-$ at their center and the false vacuum $\phi_+$ outside.

\begin{figure}[h]
\includegraphics[width=0.45\textwidth]{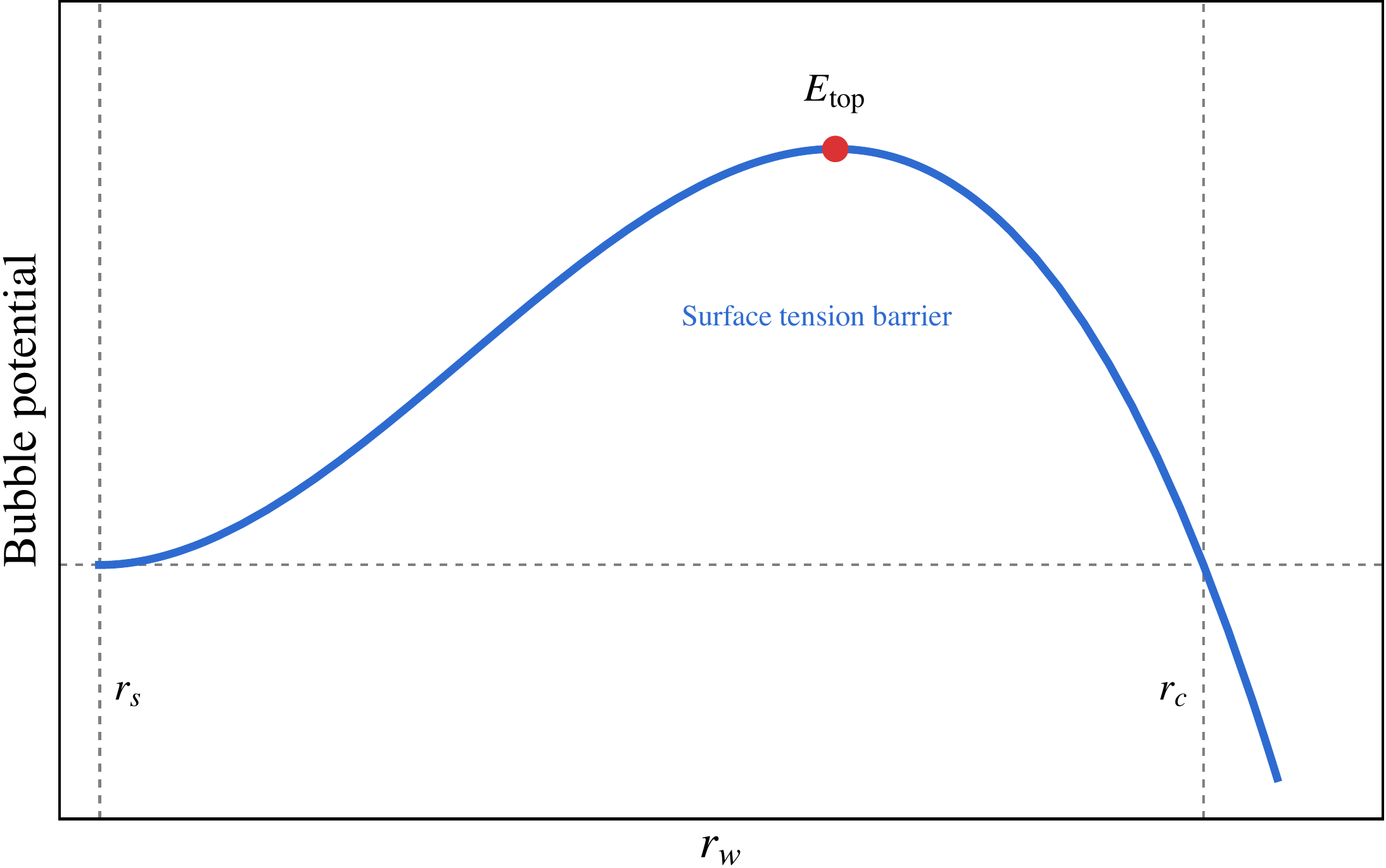}
\caption{Schematic bubble potential as a function of the bubble radius $r_w$ for a bubble centered around a black hole with Schwarzschild radius $r_s$. The barrier from the horizon to the exterior is the surface-tension barrier, and the energy at the maximum of the potential is $E_{\rm top}$. The schematic potential is computed from $U[r,\gamma=f_w^{1/2}]$ in Eq.~\ref{eq:Uwall}}.
\label{fig:bubble_potential}
\end{figure}

In this work, we compute the production rate of on-shell, spherically symmetric bubbles around the black hole. Working in the thin-wall limit, the evolution of these bubbles is controlled by the interplay between a bulk pressure term and a surface tension term. The latter provides a potential barrier for the radial growth of bubbles. 

In this paper we study in detail the evolution of boosted bubbles as they traverse their surface-tension barrier, illustrated in Fig.~\ref{fig:bubble_potential}. We focus on small enough black holes $r_s\ll r_c$, where the generation process of on-shell bubbles factorizes into four distinct stages: (a) Hawking-production; (b) a burst of radiative losses; (c) classical/quantum motion over/through the barrier; and (d) the runaway evolution of on-shell bubbles. Our analysis focuses exclusively on spherically symmetric bubbles in the thin-wall limit. In addition, we assume that the bubble thickness is smaller than the Schwarzschild radius and that the bubble energy is much smaller than the black-hole mass,  so that  gravitational backreaction can be  neglected.

\section{Near-Horizon Kinks}\label{sec:therm}

Kinks are solutions of the equation
\begin{eqnarray}\label{eq:Kinkeq}
\phi''_{kink}(\xi)=V'(\phi_{kink}(\xi))\,,
\end{eqnarray}
with $\lim_{\xi\rightarrow\pm\infty}\phi_{kink}(\xi)=\phi_\mp$. For $\xi=\gamma(r-vt-r_0)$, $\phi_{kink}(\xi)$ are also exact solutions for the Klein-Gordon equation in flat $1+1$ dimensions. One can also define radial kink-like solutions to the KG equation in $3+1$ dimensions; these, however, are not exact solutions and can radiate at sufficiently large Lorentz boosts. 

We now consider kink solutions in the near-horizon region of a Schwarzschild black hole. As is well known, this region corresponds to Rindler space, namely, as $r\rightarrow r_s$ the Schwarzschild metric approaches
\begin{eqnarray}\label{eq:NHmetric}
ds^2=\frac{4r^2_s}{\exp(1)}\left[-dT^2+dX^2\right]+r^2_sd\Omega^2\,,
\end{eqnarray}
where $T=\tfrac{1}{2}(V+U),\,X=\tfrac{1}{2}(V-U)$ and $U,V$ are Kruskal--Szekeres coordinates,
\begin{eqnarray}\label{eq:intmap}
U=-\exp\left[\frac{r_*-t}{2r_s}\right]~,~V=\exp\left[\frac{r_*+t}{2r_s}\right]\,.
\end{eqnarray}
Here $r_*\equiv r+r_s\log[(r/r_s)-1]$ is the usual Schwarzschild tortoise coordinate. The KG equation in the metric \eqref{eq:NHmetric} has ordinary kink solutions $\phi=\phi_{kink}(\xi^{Rind})$ where $\phi_{kink}$ solves \eqref{eq:Kinkeq}. The argument of this kink is 
\begin{eqnarray}
\xi^{Rind}&=&2r_s\left[\bar{\gamma}(X-\bar{\beta} T)+e^{\frac{\bar{r}_*}{2r_s}}\right]\nonumber\\[5pt]
&=&r_s\left[e^{-\frac{\bar{t}}{2r_s}} V-e^{\frac{\bar{t}}{2r_s}} U+2\gamma_{i}\right] ~~\label{eq.kink_rindler}
\end{eqnarray}
where $\bar{\gamma}=(1-\bar{\beta}^2)^{-1/2}\equiv \cosh(\frac{\bar{t}}{2r_s})$ is the boost in $T,X$ coordinates, while $\gamma_i$ is the initial value of the boost $\gamma_w$ defined below in Section~\ref{sec:SchBW}. The conserved energy of this solution is 
\begin{eqnarray}\label{eq:kinkE}
E=4\pi\sigma \gamma_i r^2_s\,,
\end{eqnarray}
with the tension $\sigma=\int d\phi \sqrt{2V(\phi)}$. Importantly, the flat metric \eqref{eq:NHmetric} is only a truncation of the near-horizon limit of the Schwarzschild metric in $(T,X,\theta,\phi)$ coordinates, neglecting gradients. As we show in the next section, highly boosted bubble walls described by $\phi=\phi_{kink}(\xi^{Rind})$ in fact radiate and lose energy until they stabilize as thin walls in the Schwarzschild background.

For the purpose of connecting near-horizon kinks to their full Schwarzschild generalization, we express $\xi^{Rind}$ in terms of $r_*,t$,
\begin{eqnarray}\label{eq.rindler_cosh}
\xi^{Rind}&=&2r_s\,
   \left[\gamma_i-e^{\frac{r_*}{2
   r_s}} \cosh \left(\frac{t-\bar{t}}{2
   r_s}\right)\right]\,.
\end{eqnarray}
In this case the near-horizon (tortoise) position $r_{*,w}(t)$ of the kink is at 
\begin{eqnarray}\label{eq:Rindsol}
&&r_{*,w}(t)=-2r_s\log\left[\gamma^{-1}_i\cosh\left(\frac{t-\bar{t}}{2r_s}\right)\right]\,. 
\end{eqnarray}
so that $\xi^{Rind}(r_{*,w}(t),t)=0$.
\section{Hawking production}\label{sec:Hawk}
We now focus on the special case in which $V(\phi)=V_{sG}(\phi)$ where
\begin{eqnarray}\label{eq:VsG}
V_{sG}(\phi)=\frac{m^2}{r_s^2\beta^{2}_{sG}}
\left[1-\cos(r_s\beta_{sG} \phi)\right]\,,
\end{eqnarray}
is the sine-Gordon potential, adapted to the units of $3+1$ dimensions. In this case, the near-horizon theory becomes effectively a $1+1$ dimensional sine-Gordon theory. This theory has a famous dual description \cite{Coleman:1974bu,Mandelstam:1975hb}, namely, the Thirring model of interacting $1+1d$ fermions, with a coupling $\lambda=\pi\left(\frac{4 \pi}{\beta^2_{sG}}-1\right)$. Under this duality, near-horizon kinks map to near-horizon massive interacting fermions. The duality was originally studied for flat $1+1d$ space, but was also extended to $AdS_2$ \cite{Callan:1982au}, to the half-line \cite{Liguori:1997vd}, and also used for the radial direction of magnetically charged black holes \cite{Maldacena:2020skw}. Like any other particle-like state, the production rate of a Thirring fermion in the near-horizon region\footnote{More precisely, this is defined as the thermal flux at an intermediate scale $r_s\ll r_m\ll r_c$} is Hawking-like:
\begin{equation}
    \Gamma \propto e^{-\frac{E}{T_H}} \label{eq.thirring_rate}
\end{equation}
where $E$ is given by \eqref{eq:kinkE}, and $T_H=(4\pi r_s)^{-1}$ is the Hawking temperature. The duality argument can be pushed further to arbitrary periodic potentials $V(\phi)$, by Fourier transforming it and using the Abelian bosonization map
\begin{eqnarray}
\cos\left(n\beta\phi\right)\leftrightarrow \left(\psi^\dagger_L\psi_R\right)^n+{\rm h.c.}\,
\end{eqnarray} 
As a result, the KG action with a general periodic $V(\phi)$ is equivalent to the massive Thirring model + higher left-right interactions \cite{Senechal}. While these interactions could lead to 1+1d confinement of the radial fermions into composites with energies $E_n$, Hawking radiation would still produce these composites with rate $\sim\exp(-E_n/T_H)$. 

To confirm our expectations of Hawking production, we also performed a complementary semiclassical analysis of bubble generation in the thin-wall limit (see Appendix~\ref{app:semiclassical}). This analysis is complementary in the sense that, on one hand, it does not rely on fermionization, while on the other hand it neglects energy loss to scalar radiation, and so is only valid for low initial boosts. The analysis in Appendix~\ref{app:semiclassical} mirrors the semiclassical approach to the Hawking radiation of particles \cite{Parikh:1999mf}, namely, it considers the complex-time evolution of Nambu--Goto kinks, that ultimately ends at real time for $r_w\rightarrow\infty$. In this analysis, we take extra care to specify an initial condition for the Nambu--Goto analysis of the bubble generation rate,  corresponding to an \textit{Unruh} vacuum near the horizon \cite{Unruh:1976db}. Finally, we obtain the semiclassical bubble generation rate, which is just \eqref{eq.thirring_rate} evaluated at the energy of the dominant saddle of the path integral (``the bounce''). This energy is nothing but $E=E_{top}$, the energy at the top of the surface-tension barrier. We reiterate that the analysis of Appendix~\ref{app:semiclassical} is only reliable when the energy loss rate to scalar radiation is negligible -- see the next section for the consideration of this crucial effect.

\section{Bubble Walls in Schwarzschild}\label{sec:SchBW}
The bubble wall/kink solution $\phi=\phi_{kink}(\xi^{Rind})$ with \eqref{eq:Kinkeq} is only valid in the near-horizon region, where the metric is approximately flat. Here we investigate the dynamics of a bubble wall beyond its near-horizon truncation. In this case we look for solutions for the (radial) Klein-Gordon equation in the full Schwarzschild metric,
\begin{eqnarray}\label{eq:KGSch}
\square_{Sch}\,\phi-V'(\phi)=0\,,
\end{eqnarray} 
where $\square_{Sch}=-f^{-1}\partial^2_t+r^{-2}\partial_r\left(r^2f\partial_r\right)$ is the d'Alembertian in Schwarzschild coordinates. To find thin-wall solutions for \eqref{eq:KGSch}, we set up \textit{Gaussian normal coordinates} to the wall in the following manner. The worldsheet of this wall is parametrized by $x^\mu_w(\tau)=\left(t_w(\tau),r_w(\tau)\right)$, where we suppress the angular directions. The velocity of the wall is $u^\mu_w=\dot{x}^\mu_w$, where the dot is with respect to the wall's proper time $\tau$. The latter is chosen so that $u^2_w=-1$. The normal to the wall is $n^\mu_w=(f^{-1}_w\dot{r}_w,f_w\dot{t}_w)$ where $f_w=1-r_s/r_w$ and $n^2_w=1,n_w\cdot u_w=0$. We also define the wall's Schwarzschild ``boost factor'' $\gamma_w\equiv n^r=\sqrt{f_w+\dot{r}^2_w}$ and its proper acceleration normal to itself, $a_w\equiv n^\mu_w u^\nu_w \partial_\nu u^w_\mu=(\ddot{r}_w+\tfrac{1}{2}f'_w)/\gamma_w$.

In the vicinity of the wall, we work in Gaussian normal coordinates (see, e.g. \cite{Blanco-Pillado:2024bev}) $(\tau,\xi)$, defined so that for any point $x^\mu=(t,r)$ we have
\begin{eqnarray}\label{eq:map}
x^\mu=x^\mu_w(\tau)+n^\mu_w(\tau)\,\xi\,.
\end{eqnarray}
The Schwarzschild metric in $(\tau,\xi)$ coordinates is 
\begin{eqnarray}\label{eq:SchGNC}
g_{\tau\xi}={\rm diag}\left(-\mathcal{Z}^2,1,r^2,r^2\sin^2\theta\right)+\mathcal{O}(\xi^2)\,.
\end{eqnarray}
where $\mathcal{Z}(\tau,\xi)=1+a_w(\tau)\,\xi$ and $r(\tau,\xi)=r_w(\tau)+\gamma_w(\tau)\xi$. The trace of the extrinsic curvature of constant $\xi$ slices is defined as $\mathcal{K}(\tau,\xi)\equiv\partial_\xi\log \sqrt{-g_{\tau\xi}}$. Explicitly,
\begin{eqnarray}\label{eq:KK}
&&\mathcal{K}(\tau,\xi)=\left(a_w+2\frac{ \gamma_w}{r_w}\right)-\left(a^2_w+2\frac{ \gamma^2_w}{r^2_w}\right) \xi+\nonumber\\[5pt]
&&\left[a^3_w+2\frac{ \gamma^3_w}{r^3_w}+\frac{r_s}{r_w^3}\left(a_w-\frac{ \gamma_w}{r_w}\right)\right] \xi^2+O\left(\xi^3\right) \,.
\end{eqnarray}
Note that $\mathcal{K}(\tau,0)$ is the trace of the extrinsic curvature of the wall itself. 

We now consider rigid wall solutions to the Klein-Gordon equation in $(\tau,\xi)$ coordinates. The latter is given by
\begin{eqnarray}\label{eq:KGGauss}
\square_{\tau\xi}\,\phi-V'(\phi)=0\,.
\end{eqnarray}
where $\square_{\tau\xi}=D_\mu D^\mu$ is the Schwarzschild d'Alembertian in $(\tau,\xi)$ coordinates.
Specifically, we look for solutions of the form $\phi=\phi_{kink}(\xi)$ solving \eqref{eq:Kinkeq}.
To check the validity of these solutions, we substitute into \eqref{eq:KGGauss} the ansatz $\phi=\phi_{kink}(\xi)+\delta\phi(\tau,\xi)$ and expand to leading order in the fluctuation $\delta\phi$. We get
\begin{eqnarray}\label{eq:KGGaussfluc}
\left[\square_{\tau\xi}-V''(\phi_{kink})\right]\delta \phi= \mathcal{K}\phi_{kink}^{\prime}\,.
\end{eqnarray}
To get an ODE for the wall's position $r_w(\tau)$, we project \eqref{eq:KGGaussfluc} on the translational zero mode $\phi^{\prime}_{kink}$:
\begin{eqnarray}\label{eq:proj}
0=\int d\xi\,\, \phi^{\prime}_{kink}\left[\square_{\tau\xi}-V''(\phi_{kink})\right]\delta \phi=\int d\xi\,\, \mathcal{K}\left(\phi_{kink}^{\prime}\right)^2~\,
\end{eqnarray}
where the first equality stems from acting to the left on $\phi'_{kink}$ -- c.f. the derivative of \eqref{eq:Kinkeq} with respect to $\xi$. From \eqref{eq:Kinkeq}, we also know that $\phi_{kink}^{\prime}$ is peaked around $\xi=0$ with height $\Delta\phi$ with width $\delta=m^{-1}$ and $m^2=V''(\phi_\pm)$. Consequently, we have
\begin{eqnarray}\label{eq:projNG}
a_w+\frac{2\gamma_w}{r_w}=\mathcal{K}(\tau,0)=m\,\mathcal{O}\left(\zeta^3_w\right)\,,
\end{eqnarray}
where $\zeta_w\equiv \frac{\delta \gamma_w}{r_w}$. The leading condition of vanishing $\mathcal{K}(\tau,0)$ is exactly the Nambu--Goto equation of motion for the wall (with zero bias $\delta V$ for now). In other words, $\mathcal{K}(\tau,0)=0$ is equivalent to $\dot{E}_{NG}=0$, where 
\begin{eqnarray}\label{eq:ENG}
\dot{E}_{NG}=\frac{d}{d\tau}\left[4\pi\sigma\gamma_wr^2_w\right]=0\,. 
\end{eqnarray}

We see that the backreaction from the fluctuation leads to corrections of order $\zeta^3_w$ to the EOM. Thus $E_{NG}$ is conserved up to perturbative corrections provided that $\zeta_w\ll 1$. For $r\rightarrow r_s$, the solution for \eqref{eq:ENG} coincides with the Rindler solution \eqref{eq:Rindsol}. Consequently, for $\zeta_w\lesssim 1$, bubbles are generated in the near horizon as Hawking radiation, and subsequently evolve according to \eqref{eq:ENG}. 

\section{Radiation from boosted vacuum bubbles}\label{sec:evol}

We now study the energy flux into scalar radiation generated by a near-horizon kink.  
For $\zeta_w\lesssim 1$, we can estimate this flux analytically, while for $\zeta_w> 1$ perturbation theory breaks down and we rely on a full numerical solution of the Klein-Gordon PDE.

\begin{figure}[ht]
  \centering
  \includegraphics[width=0.45\textwidth]{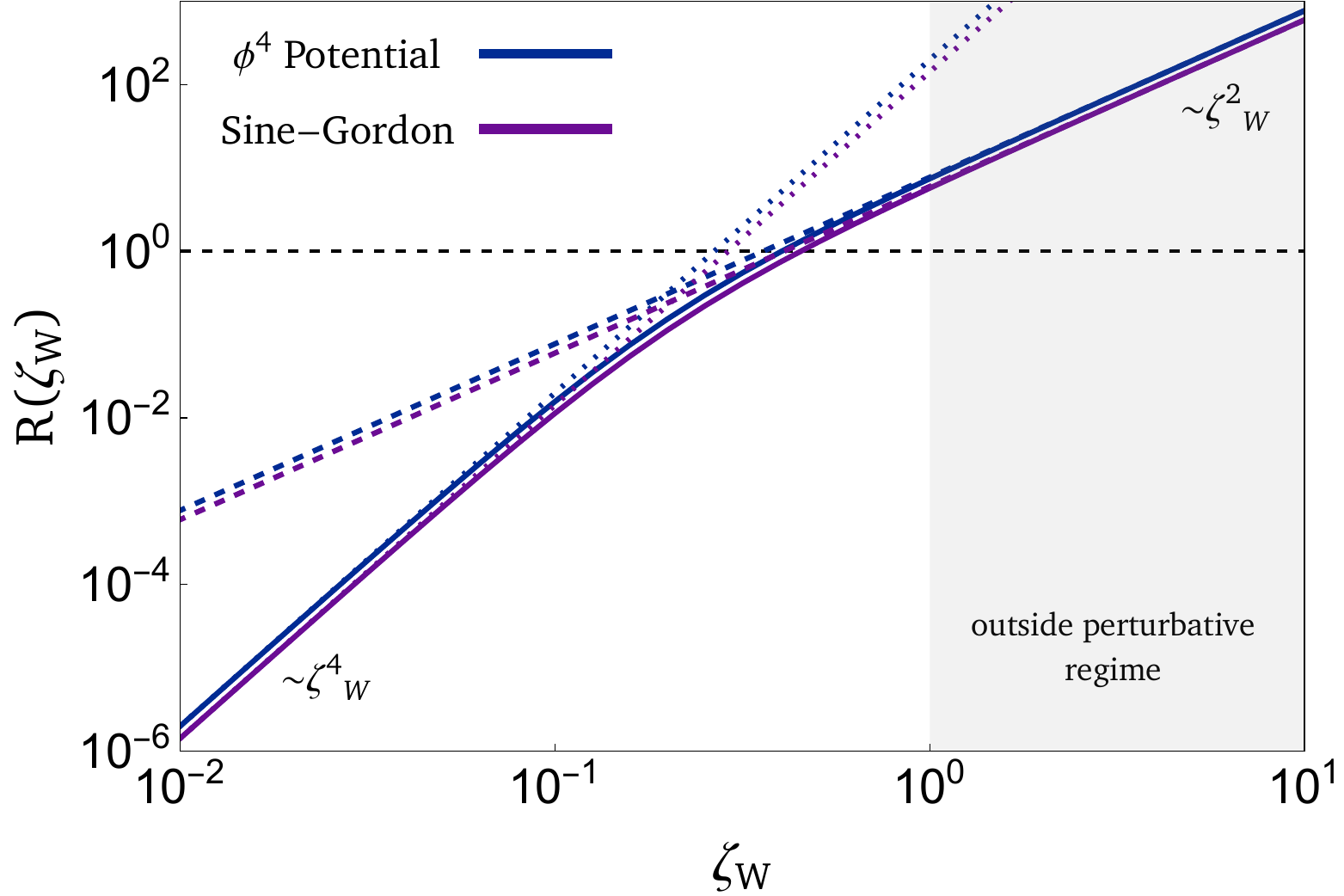}
  \caption{
  Emission rate $R(\zeta_w)$ (normalized by the inverse crossing time $r^{-1}_s$) for the $\phi^4$ model (blue) and the sine-Gordon model (purple). The dotted lines are at $(\zeta_w/\zeta_{IR})^{4}$ and the dashed lines are at $(\zeta_w/\zeta_{UV})^{2}$. The horizontal dashed line is at $R=1$.
  }
  \label{fig:Rzeta}
\end{figure}

To estimate the rate for $\zeta_w\lesssim 1$, we consider the trace of the extrinsic curvature \eqref{eq:KK} evaluated on the Nambu--Goto solution, namely $a_w=-2\gamma_w/r_w$. We have
\begin{eqnarray}\label{eq:mapnew}
&&\mathcal{K}(\tau,\xi)|_{NG}=-\frac{3}{2}a^2_w(\tau)\xi+O\left(\xi^2\right) \,.
\end{eqnarray}
We also greatly simplify our estimate by taking the near-horizon limit, in which $\square_{\tau\xi}=-\partial^2_\tau+\partial^2_\xi$. In this case the emitted power is given by
\begin{eqnarray}\label{eq:loss}
\dot{E}_{\mathrm{loss}}(\tau)&=&4\pi r^2_s\left(\frac{3}{2}\right)^2\int_{-\infty}^\tau d \tau'  a_w^2(\tau)a_w^2(\tau')\times\nonumber\\[5pt]
&& \int \frac{d k}{2\pi}\,|I(k)|^2 \cos \left[\omega_k(\tau-\tau')\right]
\end{eqnarray}
where 
\begin{eqnarray}
I(k)=\int d\xi \,\xi\,\phi'_{kink}(\xi)e^{-i\xi k}=i\partial_k\tilde{\phi}_{kink}(k)\,,
\end{eqnarray}
is the Fourier transform of the source, $\tilde{\phi}_{kink}(k)$ is the Fourier transform of $\phi_{kink}(\xi)$ , and $\omega_k=\sqrt{k^2+m^2}$. 
The $\tau'$ integral in \eqref{eq:loss} is dominated by its upper limit, and $a_w(\tau')$ on the Nambu--Goto solution is approximately $a_w(\tau')=a_w(\tau)\exp\left[3m\zeta_w(\tau'-\tau)\right]$ for $\tau^\prime\approx\tau$. Substituting this into \eqref{eq:loss} and performing the $\tau'$ integral, we get the normalized emission rate $R\equiv r_s\frac{\dot{E}_{\mathrm{loss}}}{E_{kink}}$,
\begin{eqnarray}\label{eq:loss2}
&&R=36\,\zeta_w^3\frac{m^3}{\sigma}\int \frac{d k}{2\pi}\,|I(k)|^2 \frac{6m\zeta_w}{k^2+(6m)^2\zeta_w^2+m^2}\,.~~~~~~
\end{eqnarray}
We normalized the rate by the inverse of the typical crossing time, which at $r_w\sim r_s$ is simply $r^{-1}_s$. The rate $R$ is shown in Figure~\ref{fig:Rzeta} for the $\phi^4$ model with $V(\phi)=\frac{\lambda}{4}\left(\phi^2-\frac{m^2}{2\lambda}\right)^2$, and the sine-Gordon model \eqref{eq:VsG}. We find that its features are universal,
\begin{eqnarray}\label{eq:Ffeat}
R(\zeta_w)=\begin{cases} \left(\frac{\zeta_w}{\zeta^i_{IR}}\right)^4 & \zeta_w\lesssim1 \\
\left(\frac{\zeta_w}{\zeta^i_{UV}}\right)^2 & \zeta_w\gtrsim1\,
\end{cases}\,.
\end{eqnarray}
Explicitly, $\zeta^{\phi^4}_{IR}=0.27$, $\zeta^{\phi^4}_{UV}=0.36$ and $\zeta^{sG}_{IR}=0.29$, $\zeta^{sG}_{UV}=0.4$. Note that $\zeta_w\gtrsim1$ is outside the validity of our perturbative computation, and we present it merely to guide the eye. Importantly, $R(\zeta_w=1)\sim1$, namely we lose perturbative control exactly when the emission rate is faster than the inverse crossing time. Consequently, bubbles generated with $\zeta_w\gtrsim1$ lose their energy promptly, as we checked by numerically solving the full Klein-Gordon PDE below. 


For $\zeta_w\sim1$, the perturbative corrections to $\phi_{kink}$ no longer converge, and so it is no longer a viable approximate solution to \eqref{eq:KGSch}. In this case, as we show numerically below, kink-like initial states radiate significantly, reducing the boost factor until $\zeta_w\lesssim 1$. We demonstrate this through an explicit solution of the radial KG equation in Schwarzschild \footnote{We also explicitly crosschecked our results with a flat space simulation which gives similar results.} with a $\phi^4$ potential and with $\delta V=0$ (when the bubble emerges from the horizon, the volume pressure term is completely negligible). The initial configuration is given by Eq.~\ref{eq.kink_rindler} in the near-horizon region. The details of the simulations are given in Appendix~\ref{app.sim_details}. 

In Fig.~\ref{fig:trajectory-gamma} we plot $r^w_{*}(t)$ for the Schwarzschild kink. We see that for $\zeta_w\gg 1$, the prompt energy-loss burst drives the wall to a nearly universal post-burst state with $\zeta_w\sim 1$, or equivalently, $\gamma_w\sim \gamma_{\oplus}\equiv \frac{r_s}{\delta} $, so that increasing the initial boost $\gamma_i$ no longer changes the subsequent conservative evolution or the maximal radius. To further test our hypothesis, in Fig.~\ref{fig:trajectory-scaled} we vary $\delta/r_s$ while taking boosts deep in the regime $\zeta_w\gg 1$. We expect that once the kink exits the near horizon region, i.e. $r_*\sim r_s$, for large enough boosts $\gamma_i> \gamma^{\oplus}$, the radiation instantaneously lowers the boost to $\gamma^{\oplus}$. Consistently with our estimate \eqref{eq:loss2}, this happens fast enough so that $r_w$ is still of order $r_s$ at the end of the burst, and so the maximal value of $r_w$ saturates at a radius denoted as $r^\oplus_{max}$ which scales with $r_s \sqrt{\frac{r_s}{\delta }} $ for all $\gamma_i\gtrsim \gamma^\oplus$. We validate our prediction for $r^{\oplus}_{\max}$ in Fig.~\ref{fig:trajectory-scaled}, where the proportionality factor is fitted to the data and found to be 1.6. 
\begin{figure}[t]
  \centering
 \includegraphics[width=0.48\textwidth]{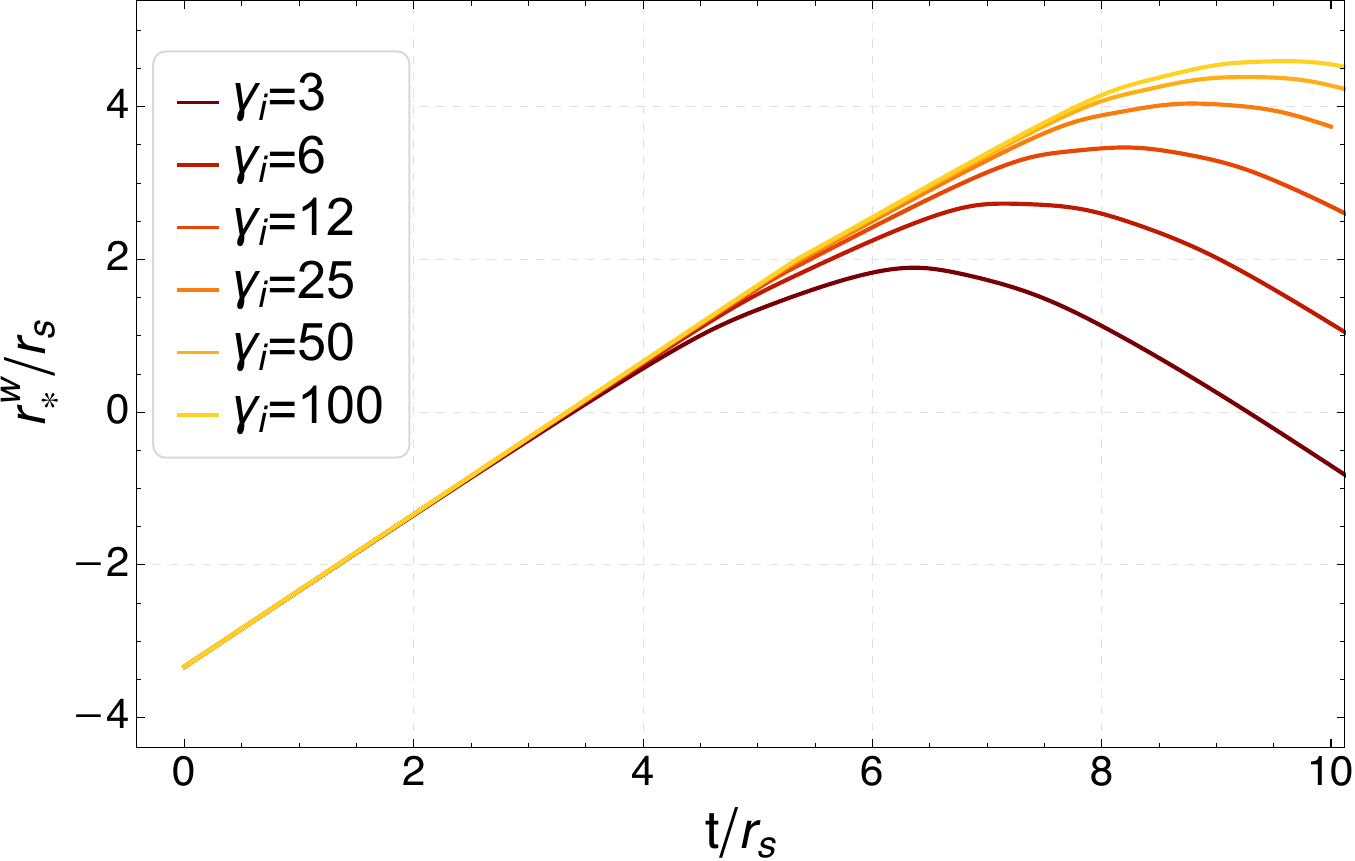}  
  \caption{
  Boosted domain-wall trajectories in Schwarzschild tortoise coordinate $r_*^w(t)$, where $r_*,t$ are given in units of $r_s$ ($r_s=6$ in the runs).  We observe that as $\gamma_i$ increases, the maximal radius $r^\oplus_{*\max}$ saturates and does not increase further. This behavior is explained by the increasing energy loss via radiation at higher boost factors. The initial state is a boosted Rindler kink defined in \eqref{eq.kink_rindler} with the initial position deep in the near horizon region $r_*(0)=-3.33 r_s$.  }
  \label{fig:trajectory-gamma}
\end{figure}
\begin{figure}[h]
  \centering
  \includegraphics[width=\linewidth]{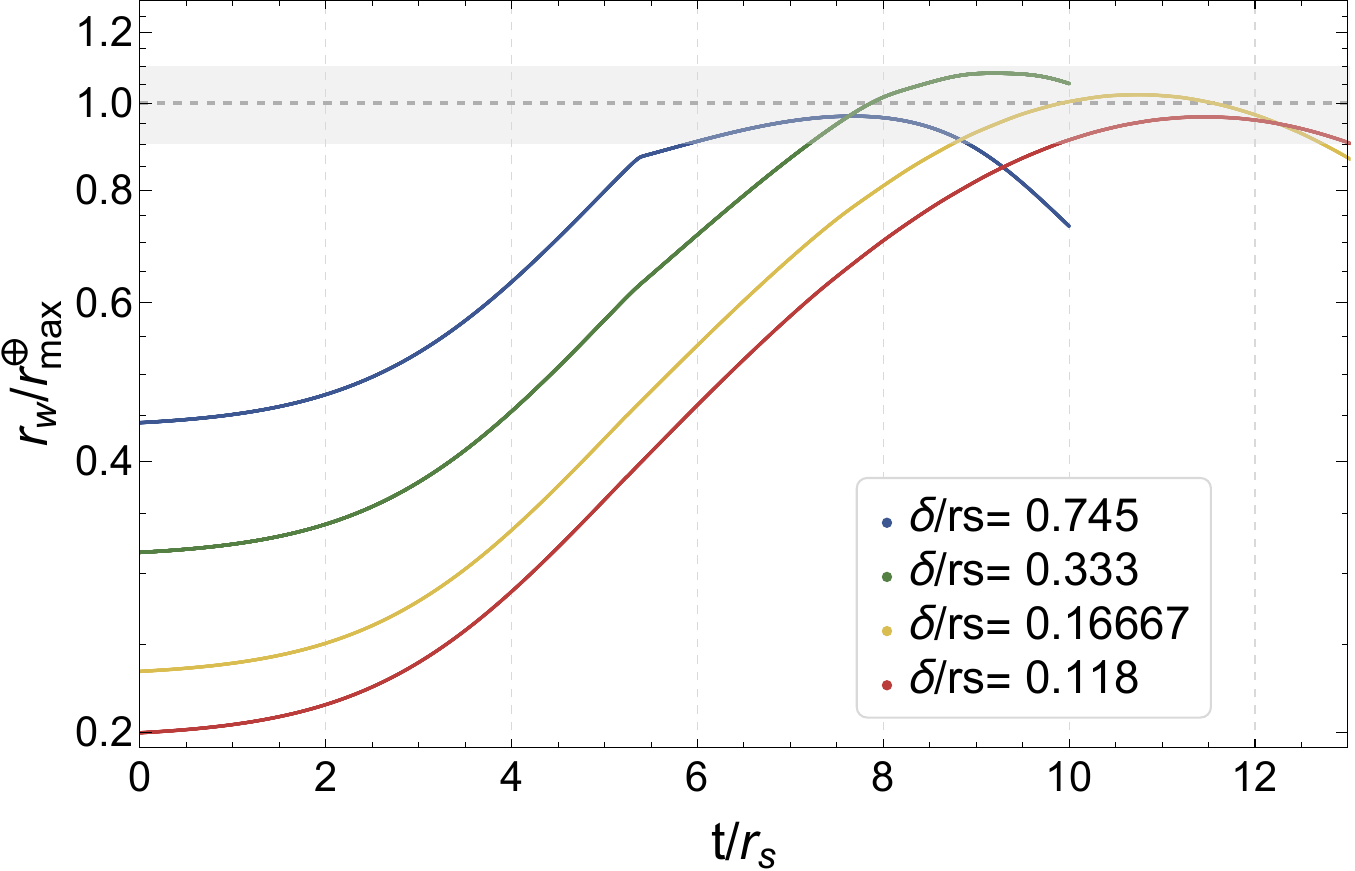}
  \caption{Boosted wall trajectories $r(t)$ for various values of the wall thickness $\delta=m^{-1}$ in units of $r_s$ and for very large initial boosts $\gamma_i\gg r_s/\delta$, where $r_{w}$ normalized by the predicted $r^{\oplus}_{\max}$, with a prefactor of $1.6$ chosen to best fit the data, and time is given in units of $r_s$. The initial state and $r_s$ are the same as in Fig.~\ref{fig:trajectory-gamma}. We find good agreement with this prediction at the $\pm10\%$ level (see the shaded band), for boosts high enough to reach saturation of $r_{\max}$ (see Fig.~\ref{fig:trajectory-gamma}). } 
  \label{fig:trajectory-scaled}
\end{figure}
\section{Estimated Tunneling rate}

In Section~\ref{sec:Hawk} we showed that the production of a bubble in the near-horizon region is consistent with a Hawking radiation process. Naively, this would mean that a sufficiently small black hole will produce bubbles with enough energy to overcome the surface tension barrier and reach infinity, with no exponential suppression. In Section~\ref{sec:evol}, however, we showed that $r^{\oplus}_{max}$ for the bubbles saturates due to radiative energy losses, and so it is so far unclear whether smaller black holes with higher $T_H$ are indeed efficient in generating runaway bubbles. In this section we show that smaller black holes are not particularly efficient at generating runaway bubbles. In fact, below a certain value of $r_s$, the production rate for runaway bubbles saturates.

We now estimate the overall rate for on-shell bubble generation. For each energy, the exponential suppression of the production amplitude is 
\begin{eqnarray}\label{eq:amp}
&&f_{prod}(E_i)
\,\mathcal{T}_{barrier}(E_\oplus[E_i])\,.
\end{eqnarray}
The first factor
\begin{eqnarray}
&&f_{prod}(E_i)= e^{-E_i/2T_H}\,,
\end{eqnarray}
is simply the amplitude for the Hawking production of a bubble wall with energy $E_i$, generated predominantly at $r_i \sim r_s$. According to the previous section, bubbles generated with large enough $E_i$ lose their energy quickly until their energy becomes 
\begin{equation}\label{eq:Emax}
   E^{\max}_{\oplus}\approx 4\pi \sigma \delta^{-1}r_s^3\,,
\end{equation}
For this reason, their subsequent evolution is determined by the energy 
$E_{\oplus}$, defined as
\begin{equation}
    E_\oplus\left[E_i\right]={\rm min}\left(E_{i},E^{\max}_{\oplus}\right)\,.
\end{equation}
 The transmission coefficient $\mathcal{T}_{barrier}$ corresponds to the tunneling rate through the surface-tension barrier. To compute it, we make use of the Nambu--Goto action for the bubble, valid in our thin-wall limit,
\begin{eqnarray}\label{eq:NG}
S_{NG}=\int d t\,\left[-4 \pi \sigma  f_w\gamma^{-1}_wr_w^2+\overline{\delta V} r_w^3\right]\,,
\end{eqnarray}
where $\overline{\delta V}=\frac{4 \pi}{3}\delta V$, and $\gamma_w=f^{1/2}_w(1-f^{-2}_w(dr_w/dt)^2)^{-1/2}$ is the same boost as before, only written in terms of a time derivative. From here and below, we solve for $r_w(t)$ as a function of Schwarzschild time $t$ rather than proper time, as is natural for a Nambu--Goto analysis. The tunneling rate through the surface tension barrier is computed from the momentum $p(E_\oplus)$ conjugate to $r_w(t)$. Computing this momentum from \eqref{eq:NG}, we get
\begin{eqnarray}\label{eq:peq}
p(E_\oplus)=f^{-1}_w\sqrt{\left[E_\oplus+\overline{\delta V}r^3_w\right]^2-f_w(4\pi\sigma r^2_w)^2}\,.
\end{eqnarray}
Correspondingly, the tunneling factor is\footnote{$\mathcal{T}_{barrier}$ itself is defined as $\exp\left[i\,\int_{r_s}^{\infty }p(E_\oplus)\,dr\right]$ where the phase is not physically relevant}  
\begin{eqnarray}
\left|\mathcal{T}_{barrier}\right|= \begin{cases}
\exp\left[-{\rm Im}\int_{r_<}^{r_>}p(E_\oplus)\,dr\right] & E_{\oplus}<E_{top}\\
1& E_{\oplus}>E_{top}
\end{cases}\,,\nonumber\\
\end{eqnarray}
where $r_{<}(E_\oplus)$ and $r_{>}(E_\oplus)$ are the classical turning points of the barrier for energy $E_\oplus$, and 
\begin{eqnarray}\label{eq:Etop}
E_{top}\approx \frac{16\pi}{27}\sigma r^2_c\,,
\end{eqnarray}
is the energy at the top of the barrier, with $r_c=r_>(E_\oplus=0)$ . To estimate the overall rate (to exponential accuracy), we can estimate:
\begin{equation}
\Gamma(t) \sim 
\left|\int_0^{E_{\oplus}^{\max}} dE_i\, e^{i E_i t}f_{prod}(E_i) 
\,\mathcal{T}_{barrier}(E_i)\right|^2
\end{equation}
when $E_{\oplus}^{\max}> E_{top}$, we can estimate this integral with the saddle point approximation where the saddle point is $E_i^{\mathrm{saddle}}=E_{top}$ (see Appendix~\ref{app:semiclassical}), while for  $E_{\oplus}^{\max}< E_{top}$ the endpoint of $E_{\oplus}^{\max}$ dominates.

\begin{figure}[h]
  \centering
  \includegraphics[width=\linewidth]{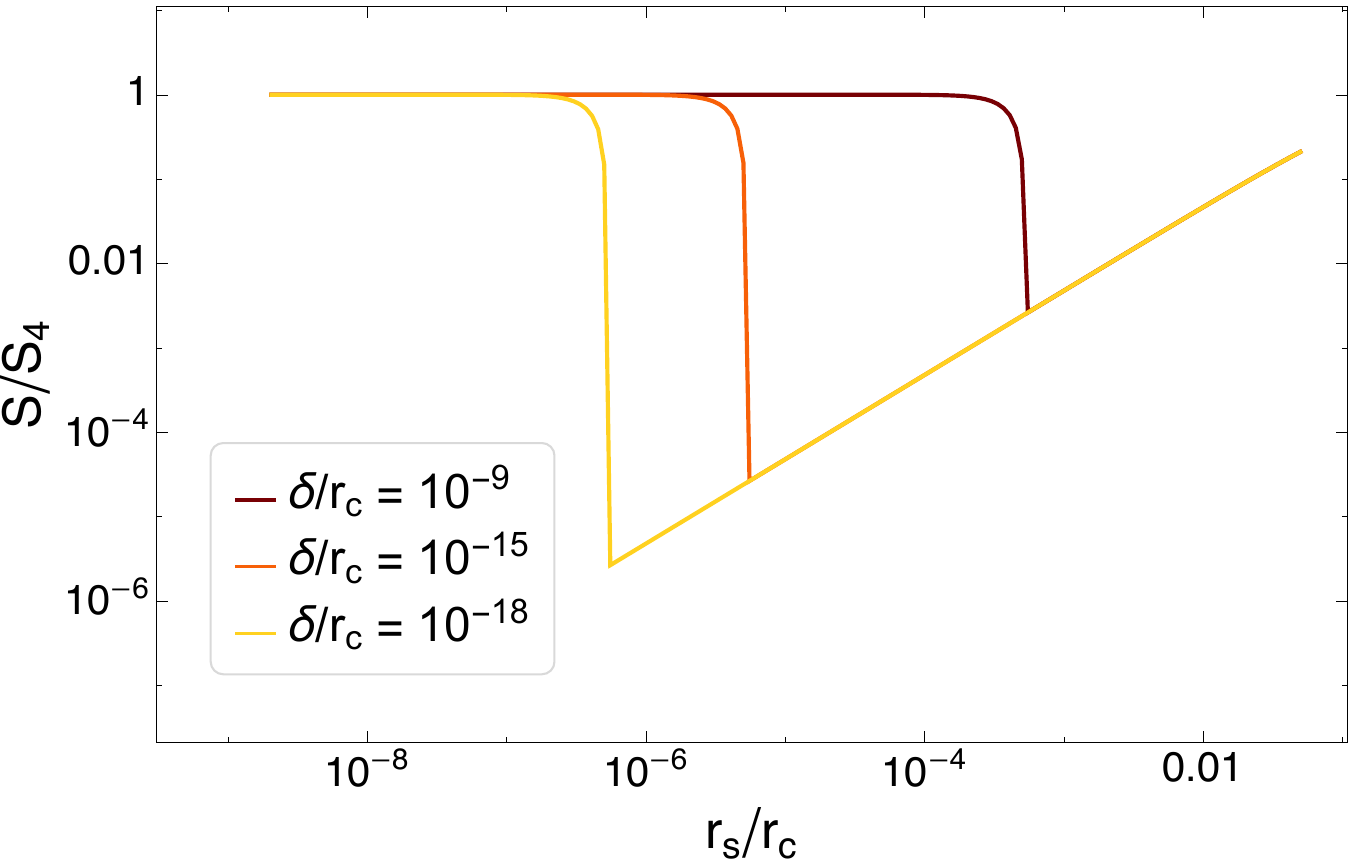}
  \caption{Full tunneling rate from \eqref{eq:tunn} normalized by the Coleman flat-space bounce action $S_4$ as a function of ratio between the Schwarzschild radius and the critical bubble radius $r_s/r_c$, for various values of the wall thickness $\delta$. }
  \label{fig:action_tunneling}
\end{figure}

Putting everything together, the tunneling rate is
\begin{eqnarray}\label{eq:tunn}
&&\log\Gamma=\nonumber\\[5pt]
&&\begin{cases}
    -T^{-1}_HE_{\oplus}-2{\rm Im}\left[\int_{r_<}^{r_>}p(E_{\oplus})dr\right]&E_{\oplus}<E_{top} \\
-T^{-1}_HE_{top}&E_{\oplus}>E_{top}
\end{cases}\,,\nonumber\\
\end{eqnarray}
The results of this computation are presented in Fig.~\ref{fig:action_tunneling}. Three regimes appear in this plot: (a) for sufficiently small $r_s$, the energy of the kink after radiation is well below the surface-tension barrier, and the rate is effectively the standard Coleman rate; (b) for larger $r_s$ (while still within $r_s\ll r_c$), the Hawking-produced bubbles remain energetic enough after radiation to move classically over the surface-tension barrier; and (c) in an intermediate regime, marginally boosted bubbles still require tunneling through the surface-tension barrier, but at an enhanced rate.      

The maximal possible rate is achieved when $E_{\oplus}\sim E_{top}$, or equivalently, 
$r_s\sim (\delta)^{\frac13}r_c^{\frac23}$. 
In this case
\begin{equation}\label{eq:rateres}
  \log \Gamma \sim  -\frac{128}{81}  \left(\frac{4\delta}{r_c}\right)^{\frac{1}{3}}S_4\,,
\end{equation}
where 
\begin{equation}
  S_4=-2{\rm Im}\left[\int_{0}^{r_c}p(E_\oplus=0)dr\right]\,,
\end{equation}
is the Coleman flat-space bounce action. The rate \eqref{eq:rateres} is exponentially faster than the spontaneous quantum tunneling rate, however it is still always exponentially suppressed. This is most transparently seen for a tilted quartic double-well potential in the thin-wall regime, for which one finds parametrically

\begin{equation}
  \log \Gamma \sim -\lambda^{-1}\left(\frac{r_c}{\delta}\right)^{\frac{8}{3}} 
\end{equation}
For perturbative $\lambda \lesssim 4\pi$, we therefore find that as long as we are in the thin-wall regime $\delta\ll r_c$, the rate remains exponentially suppressed. Our thin-wall derivation of the $\lambda^{-1}$ suppression of the exponential is consistent with the thick-wall derivation of \cite{Strumia:2022jil}.

\section{Conclusions}

In this paper we studied the possibility of realistic (i.e. non-eternal) small black holes inducing false vacuum decay within the thin-wall regime of the $\phi^4$ and sine-Gordon models. In this regime, the problem can be factorized into a Hawking production in the near horizon region, a subsequent propagation and tunneling through the surface tension barrier. For very small black holes, the Hawking temperature is high one might expect that bubbles with enough energy to overcome the surface tension barrier can be produced without any exponential suppression. However, we found that such highly boosted bubbles promptly lose energy via scalar radiation. As a result, while the vacuum-decay rate can be enhanced relative to the Coleman $O(4)$ flat-space instanton, it is always exponentially suppressed. In our analysis we only considered spherical configurations and neglected gravitational backreaction and additional radiation channels. We see no immediate reason for these corrections to significantly impact our conclusions.

\section*{Acknowledgments}
We thank Giuseppe Rossi and Binyamin Vilk, for collaboration on an earlier version of the project. We give special thanks to Raman Sundrum and Misha Smolkin for their helpful comments on the draft. We thank Amit Sever, Yonit Hochberg, Eric Kuflik, Gilad Perez, Tin Suljemanpasic, Tomer Volansky, and Shimon Yankielowicz for insightful discussions. OT is supported by the ISF grant No. 3533/24, by the NSF-BSF grant No. 2022713, and by the BSF grant No. 2024169.MG is supported by the Israel Science Foundation under Grant No. 1424/23 and by the NSF-BSF grant 2023711.
\appendix
\section{The Semiclassical Approach }\label{app:semiclassical}

In this paper we motivated the Hawking-production of boosted bubbles using fermionization in the near-horizon region. To further support our argument, we provide a self-contained semiclassical derivation of the bubble generation rate. The present derivation assumes a rigid kink and so is valid for bubbles generated with low boosts for which scalar radiation is negligible.

Our semiclassical analysis is valid in the thin-wall limit, where the dynamics of the rigid bubble is captured by the Nambu--Goto action \eqref{eq:NG}. Importantly, we set up our semiclassical computation so that the initial false vacuum is an ``Unruh'' vacuum corresponding to an astrophysical black hole, rather than a ``Hartle--Hawking'' vacuum relevant for eternal black holes (see \cite{Shkerin:2021xsh,Shkerin:2021uun} for the importance of this fact). To specify an Unruh vacuum, we match our thin-wall solution to a full solution of the KG system \eqref{eq:4daction} in the near horizon region, which is continuously connected to the false vacuum prior to black hole formation. 

To compute the rate for bubble generation, we need to calculate the rate

\begin{equation}
\Gamma=|\left\langle\text{On-shell bubble state}|\mathcal{T}e^{-i H \Delta t}|\Omega_{\mathrm{false,Unruh}}\right\rangle|^2
\end{equation}
with $\Delta t\to \infty$. 
The exponential suppression of this amplitude is given by the saddle-point action as 
\begin{equation}\label{eq:Mamp}
\Gamma=\left|\int D\phi\,e^{-\mathrm{Im}[S(\phi)]}\right|^2  
\end{equation}
where $S$ is the KG action \eqref{eq:4daction} evaluated on a saddle point (``bounce'') trajectory. The boundary conditions on this path integral are false vacuum at $t\rightarrow-\infty$ and an on shell bubble at $t\rightarrow \infty$. Since we are working in the thin-wall limit, we can instead use \begin{equation}\label{eq:Mamp_NG}
\Gamma\approx\int Dr_w\,\,e^{-2\mathrm{Im}[S_{NG}(r_w)]}\,,
\end{equation}
where $S_{NG}$ is the Nambu--Goto action \eqref{eq:NG}, and the integration is over a trajectory $t(r_w)$ that solves the Euler-Lagrange equations derived from $S_{NG}$. The relevant saddle point trajectory starts at $t(r_w\rightarrow r_s)\in\mathbb{C}$ and ends at $t(r_w\rightarrow\infty)\in \mathbb{R}\rightarrow\infty$.  Note, however, that to specify our initial false vacuum, we also need to specify a boundary condition on $(t,r_w)$ as $t\in \mathbb{R}\rightarrow-\infty$. To find the right boundary condition corresponding to an Unruh vacuum, we consider near-horizon dynamics.

\subsection{Near-horizon Dynamics}
In Section~\ref{sec:therm} we described the near-horizon dynamics of Schwarzschild bubble walls by working in the Rindler \textit{truncation} of the metric near $r_s$. While this truncation was perfectly suitable for motivating the Hawking production of bubble walls using the duality to Thirring fermions, the semiclassical approach requires a slight refinement of the Rindler treatment. The reason is that in Rindler space, modes never reach $r\rightarrow\infty$. For this reason, there is no separation between left- and right-moving modes in Rindler space, and hence no Unruh vacuum \cite{Shkerin:2021xsh}. In other words, our solution $\phi_{kink}(\xi^{Rind})$ is generated in a \textit{Hartle--Hawking} vacuum \cite{Hartle:1976tp}. Nevertheless, the solution \eqref{eq.kink_rindler} allows us to fix the near-horizon behavior of the bubble in the \textit{full Schwarzschild metric} so that it starts from an Unruh vacuum as required.

Though the Rindler solution \eqref{eq.rindler_cosh} is in a Hartle--Hawking vacuum and thus not directly applicable here, it does allow us to make an educated guess of the near-horizon boundary condition for the Nambu--Goto dynamics. To this end, we propose the near-horizon field configuration $\phi=\phi_{kink}(\xi^{NH})$ where
\begin{eqnarray}\label{eq.NH_exp}
\xi^{NH}&=&2r_s\, 
   \left[\gamma_i-\frac{1}{2}e^{\frac{r_*-\left(t-\bar{t}\right)}{2
   r_s}}\right]\,.
\end{eqnarray}
We immediately see that $\xi^{NH}\rightarrow\xi^{Rind}$ as $t\rightarrow -\infty+i\,{\rm Im}(\bar{t})$. Similarly, the position of the kink \eqref{eq.NH_exp} is given by
\begin{eqnarray}\label{eq:NHsol}
&&r^{NH}_{*,w}(t)=t-\bar{t}+2r_s\log 2\gamma_i\,,
\end{eqnarray}
and so $r^{NH}_w(t)\rightarrow r^{Rind}_w(t)$ as $t\rightarrow -\infty+i\,{\rm Im}(\bar{t})$. In other words, the kink $\phi_{kink}(\xi^{NH})$ represents the correct limit of the full Schwarzschild EOM at $t\rightarrow -\infty+i\,{\rm Im}(\bar{t})$ and $r\rightarrow r_s$. Furthermore, we can choose ${\rm Im}(\bar{t})$ so that $\phi_{kink}(\xi^{NH})$ satisfies the correct Unruh boundary condition at $t\rightarrow -\infty$, and so it can be used to set the correct initial condition on the Nambu--Goto integral \eqref{eq:Mamp_NG}. Evaluating this solution at very early real times $t\to -\infty $, we can impose Unruh boundary conditions around the false vacuum (see \cite{Shkerin:2021xsh}). Close to the horizon, for $r_* \ll t-\bar{t}$,  we find that $\mathrm{Re}\left(\frac{\xi^{NH}}{\delta}\right)\gg 1$ (this is the thin wall assumption $r_s \gg \delta $ that we use throughout the paper). Consequently, this region is at the tail of the kink, where $\phi$ can be written as 
\begin{equation}
    \phi-\phi_{+} \propto e^{-m\xi_{NH}}= e^{-2m r_s \gamma_i } e^{r_s m\exp\left(\frac{r_*-t+\bar{t}}{2r_s}\right) }\,.
\end{equation}
Here $m$ is the mass term near the false vacuum.  If we take a Fourier transform of this tail we find that 
\begin{equation}
\tilde f(\omega)
=e^{-2m r_s \gamma_i }
\frac{1}{\kappa}
\left(\frac{m}{2\kappa}e^{\kappa(r_*-\bar t)}\right)^{-i\omega/\kappa}
\Gamma\!\left(i\frac{\omega}{\kappa}\right)
\end{equation}
where $\kappa =\frac{1}{2r_s}$ is the surface gravity. We immediately notice that these modes satisfy the thermal 
conditions 
\begin{equation}
\left|\frac{\tilde f(-\omega)}{\tilde f(\omega)}\right|
=
e^{-2\pi\omega/\kappa}, \label{eq.thermal_ratio}
\end{equation}
provided that we set ${\rm Im}(\bar{t})=2\pi r_s$. This solution includes only right moving modes, as expected from the Unruh vacuum. 

All-in-all, we learn from our near-horizon analysis the correct initial condition for our Nambu--Goto analysis in the following manner. Looking at \eqref{eq:NHsol}, we see that at $r_*\rightarrow-\infty$ we need ${\rm Re}(t)\rightarrow-\infty$ and ${\rm Im}(t)\rightarrow{\rm Im}(\bar{t})=2\pi r_s$. This is the correct initial condition reflecting generation from an Unruh vacuum.

\subsection{Nambu--Goto Trajectory}
We now consider solutions for the EOM of the Nambu--Goto action \eqref{eq:NG}. To do this, we define the ``potential''
\begin{eqnarray}\label{eq:Uwall}
U[r,\gamma]\equiv4\pi\sigma\gamma r^2-\overline{\delta V}r^3\,,
\end{eqnarray}
where $f(r)=1-r_s/r$. The conserved energy of the bubble is given by $E=U[r_w,\gamma_w]$ where $\gamma_w=f^{1/2}_w(1-f^{-2}_w(dr_w/dt)^2)^{-1/2}$. The EOM is then
\begin{eqnarray}\label{eq:Encon}
0=\frac{dE}{dt}=\frac{d}{dt}U[r_w,\gamma_w]\,.
\end{eqnarray}
For later reference, let us consider the point $r_{top}$, for which $\frac{d}{dr}U[r,\gamma=f_w^{1/2}]|_{r=r_{top}}=0$. For $E=E_{top}\equiv U[r_{top},\gamma=f_w^{1/2}]$, the wall reaches an unstable equilibrium at $r_w=r_{top}$ where it rests on top of its potential barrier. $E_{top}$ is very well approximated by \eqref{eq:Etop}. Following Coleman, from now on we focus on solutions with $E\approx E_{top}$. These will give the dominant contribution to the generation of on-shell bubbles.

\begin{figure}[t]
  \centering
  \begin{subfigure}{0.23\textwidth}
         \centering
         \includegraphics[width=\textwidth]{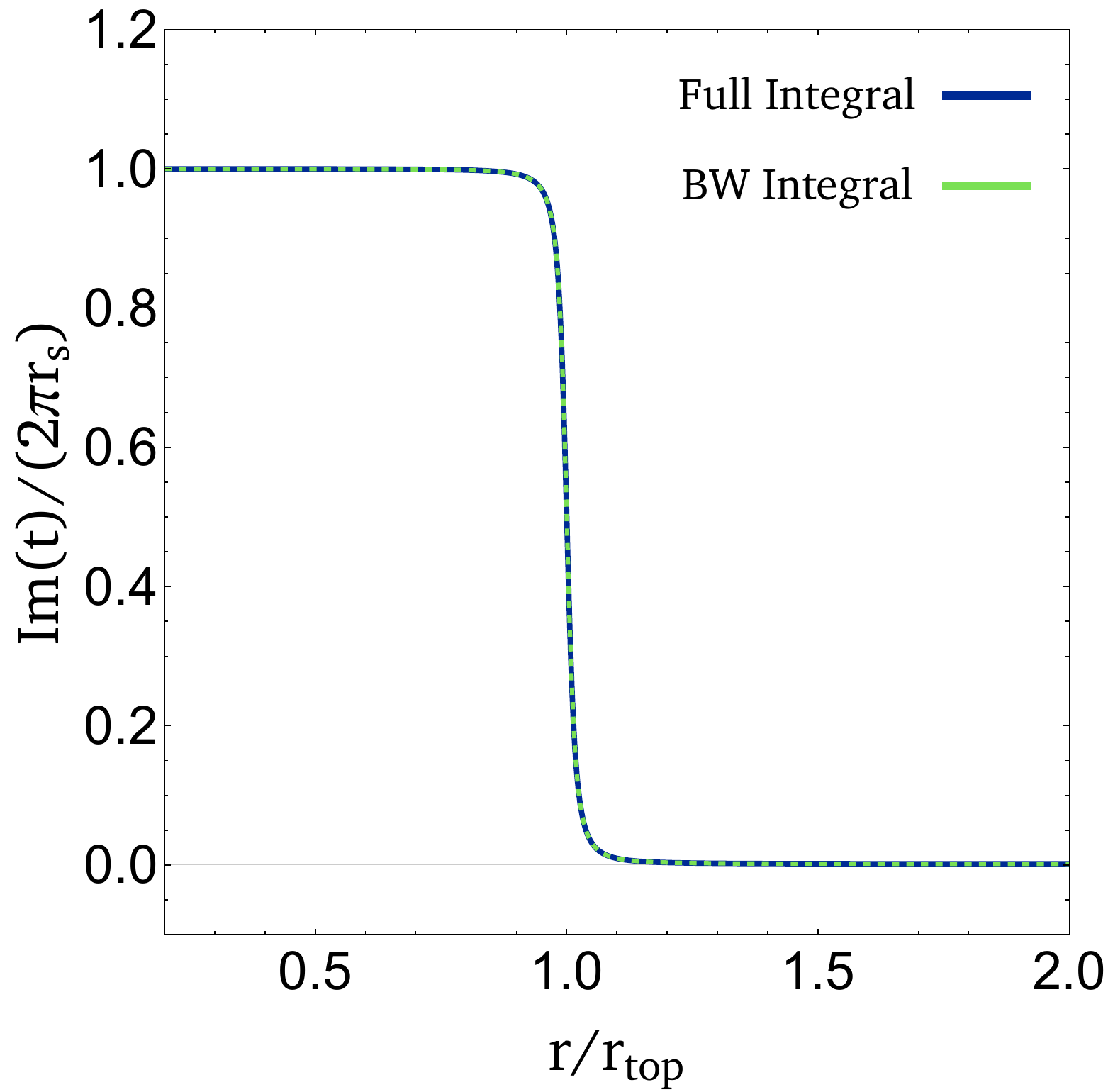}
\end{subfigure}
 \begin{subfigure}{0.23\textwidth}
         \centering \includegraphics[width=\textwidth]{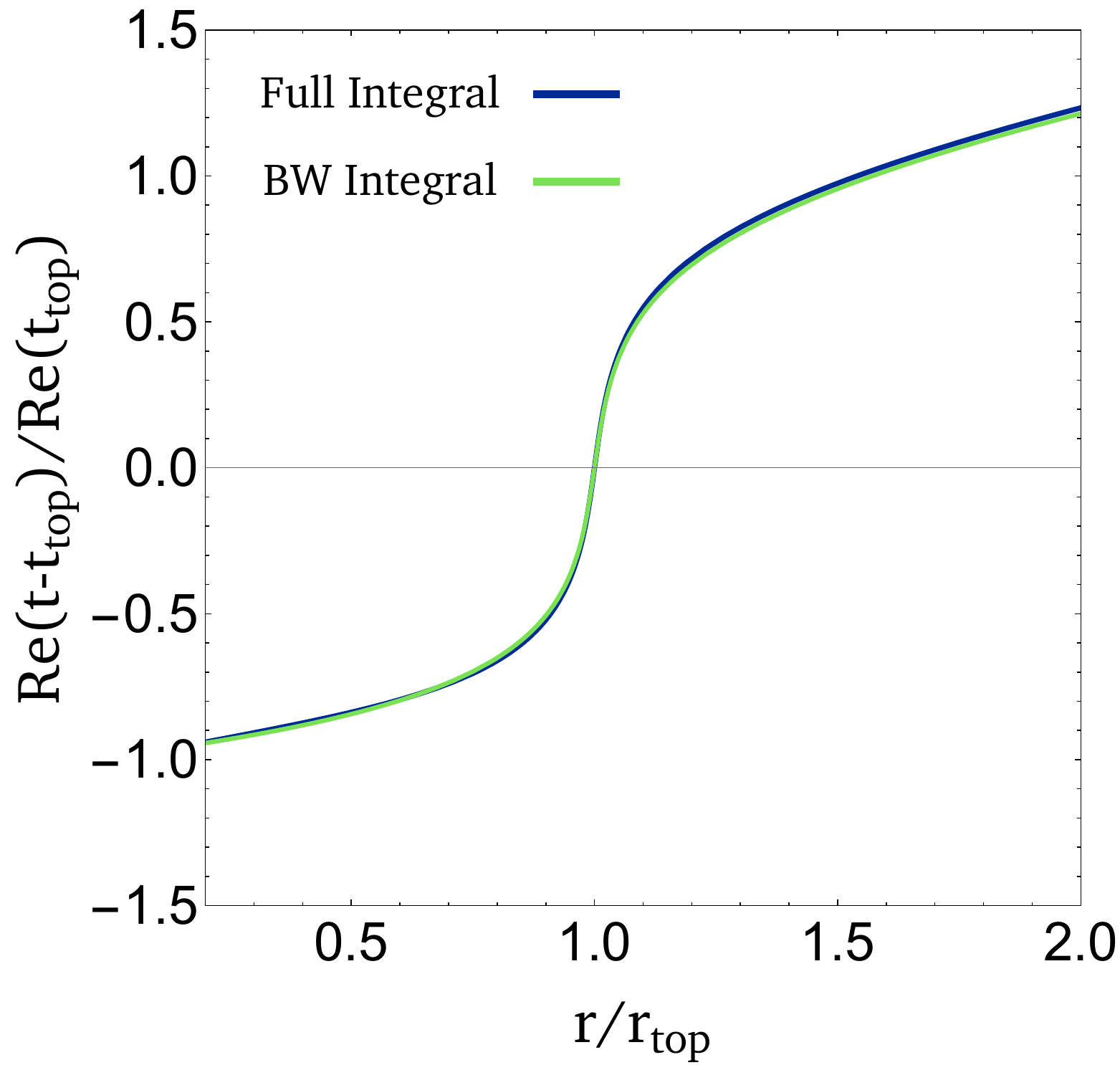}
     \end{subfigure}
     \caption{Real and imaginary parts of $t(r)$ of the bubble, using \eqref{eq:tint} (blue) and \eqref{eq:tintBW} (green). $\delta_E$ was chosen so that \eqref{eq:intIm} vanishes and the bubble ends up on-shell. Here we chose the values $r_s=2.3,\,\sigma=74$, and $\overline{\delta V}=0.1$.}\label{fig:tint}
\end{figure}

To solve \eqref{eq:Encon}, we first convert it to an ODE for $t(r)$ and integrate with respect to $t$. We get
\begin{eqnarray}\label{eq:tint}
t(r)=t_0+\int_{r_0}^r\,dr\,\frac{1}{f\sqrt{1-\left(\frac{4\pi\sigma f(r) r^2}{E+r^3\overline{\delta V}}\right)^2}}\,.
\end{eqnarray}
Here $(t_0,r_0)$ are the initial time and position for energy $E_{top}$, which we get from the near horizon solution \eqref{eq:NHsol}. Note that ${\rm Im}(t_0)={\rm Im}(\bar{t})=2\pi r_s$ in order to satisfy the Unruh boundary condition. Consider now analytically continuing by adding an infinitesimal complex part to $E$:
\begin{eqnarray}
E=E_{top}(1-\epsilon e^{i\pi\delta_E})\,,
\end{eqnarray}
where $\epsilon\ll1$ but $\delta_E$ isn't necessarily small. We are looking for trajectories so that $t(r\rightarrow\infty)$ is real, meaning that the bubble becomes asymptotically on-shell. In the thin wall limit, we can approximate the integrand of \eqref{eq:tint} exponentially well with a Breit-Wigner-like form 
\begin{eqnarray}\label{eq:tintBW}
t(r)=t_0+\int_{r_0}^r\,dr\,\frac{N_{BW}}{\sqrt{(r-r_{top})^2+\Gamma_{BW}e^{i\pi\delta_E}}}\,,
\end{eqnarray}
where $N_{BW}=f^{-1}(r_{top})\sqrt{(E_{top}+r^3_{top}\overline{\delta V})/U''(r_{top})}$ and $\Gamma_{BW}=2E_{top}\epsilon/U''(r_{top})$. Doing this integral analytically, we get
\begin{eqnarray}\label{eq:intIm}
{\rm Im}[t(r\rightarrow \infty)]=\pi \left[2r_s+N_{BW}\left(1-\delta_E\right)\right]\,,
\end{eqnarray}
So that for $\delta_E=1-\frac{2r_s}{N_{BW}}$ the bubble is generated on shell, independently of $\epsilon$, which we can take to be as small as we wish. To summarize, we found a trajectory $(t,r_w)$ that starts at complex time (due to Hawking generation) and ends at real time with an on-shell bubble at $r_w\rightarrow\infty$. See Fig.~\ref{fig:tint} for a representative trajectory, for the full integral \eqref{eq:tint}, and the Breit-Wigner-like approximation \eqref{eq:tintBW}.

\vspace{10pt}

The Nambu--Goto action evaluated on this trajectory gives the rate of on-shell bubble generation. This amplitude is given by:

\begin{eqnarray}\label{eq:rate}
\Gamma\sim\exp\left[-2{\rm Im}(S_{NG})\right]=\exp\left(-E_{top}/T_H\right)\,,
\end{eqnarray}

where $T_H=(4\pi r_s)^{-1}$. Note that to reach $r\to \infty$, we need $\epsilon\to 0$, which means that an on-shell bubble is generated with $E=E_{top}$.


Importantly, this analysis neglects radiation from the bubble and so it does not reflect the full result of this paper. It should be interpreted as a strong cross-check of the Hawking production \eqref{eq.thirring_rate} of the bubble, which is motivated in the main text via radial fermionization.

\section{Radial Schwarzschild Klein--Gordon Simulation \label{app.sim_details} }
We numerically evolve a spherically symmetric $\phi^4$ scalar field on a fixed Schwarzschild background using the tortoise coordinate $r_*$. Writing the field equation in first-order-in-time form:
\begin{eqnarray}
\partial_t\phi&=&\pi\nonumber \\
\partial_t\pi&=&\partial_{r_*}^2\phi+(2f/r)\partial_{r_*}\phi-f\,\partial_\phi V
\end{eqnarray}
with $f=1-r_s/r$ and $V(\phi)=\lambda(\phi^2-v^2)^2/4$, we discretize on a uniform $r_*$ grid with second-order centered finite differences, and integrate in time with a fourth-order Runge–Kutta scheme. The initial configuration is a boosted kink of Eq~\ref{eq.kink_rindler},
together with the consistent conjugate momentum $\pi=\partial_t\phi$. Radiative boundary conditions are imposed at the inner and outer edges to model, respectively, ingoing flux into the horizon and outgoing flux toward infinity. During the evolution we track the wall position from the rightmost $\phi=0$ crossing and monitor the total domain energy. We use 30 grid points per initial wall thickness and the timestep is determined by CFL conditions $dt=0.25dr_*$. The domain size is always taken to be large enough, and we checked for convergence with resolution. 
The code is available in the ancillary files of this submission.

\end{document}